\documentstyle[preprint,aps]{revtex}

\begin{document}
\title{Quantum Anomaly of the Transverse Ward-Takahashi Relation for the 
Axial-Vector Vartex}
\author{Han-xin He $^*$ \footnotetext{$^*$ E-mail: hxhe@iris.ciae.ac.cn}}
\address{
China Institute of Atomic Energy, P.O.Box 275(18), Beijing 102413, P.R.China
\\
and Institute of Theoretical Physics, Academia Sinica, Beijing 100080, 
P.R.China }

\maketitle
\begin{abstract}
We study the possible quantum anomaly for the transverse Ward-Takahashi 
relations in four dimensional gauge theories based on the method of 
computing the axial-vector and the vector current operator equations.
In addition to the well-known anomalous axial-vector divergence equation 
(the Adler-Bell-Jackiw anomaly), we find the anomalous axial-vector curl 
equation, which leads to the quantum anomaly
of the transverse Ward-Takahashi relation for the axial-vector 
vertex. The computation shows that there is no anomaly for the transverse
Ward-Takahashi relation for the vector vertex.


\end{abstract}

\newpage

In quantum field theory symmetries lead to relations 
among the Green functions of the theory, which are generally known as the Ward
-Takahashi(WT) identities[1]. They play an important role in providing a 
consistency condition in the perturbative 
approach as well as in the nonperturbative approach of any quantum field theory. 
In this regard, an very interesting problem 
relates to the nonperturbative studies of gauge field theories using the Dyson
-Schwinger equation(DSE) approach[2][3].
The DSEs are an infinite set of
coupled integral equations which relate the n-point Green function to the 
(n+1)-point function; at its simplest, propagators are related to three-point
vertices and so on. Therefore, we must find some way to truncate this set of
equations. If we can express the
three-point vertices in terms of the propagators, these equations will 
form a closed system for the propagators.
Naturally, the basic approach of solving this problem is to use the WT 
identities. But the normal WT identities specify only the longitudinal parts 
of Green functions, leaving the transverse parts undetermined[4].
How to solve exactly the transverse parts of vertices and so the full
vertex functions then become a crucial problem [3][4]. 
Obviously the key point is
to study the constraint on the transverse parts of vertices imposed by the
symmetry of the system, i.e. the transverse WT relations, as in the case of
the longitudinal parts of vertices.
In Ref.[5], we studied the transverse WT relation for the vector vertex. 
It showed that in order to obtain the complete solution for the vector vertex 
we need to build simultaneously the WT relations for the axial-vector and 
the tensor vertices. Up to the effects of quantum anomaly, this problem was 
solved by He[6] who presented the transverse WT relations for the vector, the 
axial-vector and the tensor vertices in four dimensional gauge
theories, and found that, indeed, the full vector and the full axial-vector 
vertex functions are expressed only in terms of the fermion propagators
in the chiral limit with massless fermions.

However, the symmetry in the classical theory may be destroied in quantum 
theory by the quantum anomaly arising from the loop diagram. The well-known 
example is the anomalous nonconservation of the four-dimensional axial current,
expressed by the anomalous axial-vector divergence equation, 
which is known as the Adler-Bell-Jackiw(ABJ) anomaly [7]. Hence the 
normal WT identity for the axial-vector vertex is modified by the ABJ anomaly. 
Therefore, in order to obtain the universally satisfied transverse WT 
relations and the full vertex functions, we need to take into account the 
quantum anomaly for them. 

In this Letter, we study the possible
quantum anomaly for the transverse WT relations
based on the method of computing the axial-vector and the vector current 
operator equations[8], respectively. In addition to the well-known anomalous 
axial-vector divergence equation, we find the anomalous axial-vector curl
equation, which leads to the quantum anomaly in the transverse WT relation 
for the axial-vector vertex. The computation shows that there is no anomaly
for the transverse WT relation for the vector vertex.

In a canonical quantum field theory, 
the transverse WT relation for the axial-vector vertex 
is related to the curl of the time-ordered products of the three-point 
function involving the axial-vector
current operator[5][6]. In order to perform the curl operation, it is 
convenient to introduce the bilinear covariant current operator,
$V_5^{\lambda \mu \nu }(x)=\frac 12\bar{\psi}(x)
[\gamma ^{\lambda} ,\sigma ^{\mu \nu }]\gamma _5\psi (x)
=i[g^{\lambda \mu }j_5^\nu (x) -g^{\lambda \nu }j_5^{\mu}(x)]$, where
$j^{\mu}_{5}(x)=\bar{\psi}(x)\gamma ^{\mu} \gamma_{5} \psi (x)$.
Thus the curl of the axial-vector current is given by 
$\partial _\lambda ^x V_5^{\lambda \mu \nu }(x)$,
where $\partial _\lambda ^x$ denotes the derivative operator with respect to 
the argument $x$.
To study the possible quantum anomaly for the transverse WT relation 
for the axial-vector vertex, let us study the operator equation
for the curl of the axial-vector current following the method of
deriving the anomalous axial-vector divergence equation[8]. 
To this end, we define the 
bilinear current operator by placing the two fermion fields at distinct 
points separated by a small distance $\epsilon$ and then carefully taking the 
limit as the two fields approach each other. Explicitly, we define
\begin{equation}
V^{\lambda \mu \nu}_5(x)=Symm {\lim_{\epsilon \rightarrow 0 }}
\{ \bar{\psi}(x+ \epsilon /2)\frac{1}{2} [\gamma ^{\lambda},\sigma^{\mu \nu}]
\gamma_5 U_P(x+ \epsilon /2, x - \epsilon /2) \psi (x - \epsilon /2)\} . 
\end{equation}
Here the Wilson line
$U_P (x^{\prime },x)=P\exp (-ig\int_x^{x^{\prime }}dy^\rho A_\rho (y))$
is introduced in order that the operator be locally gauge invariant,
where $A_\mu $ is the gauge fields.
 In the QED case, $g = e$ and $A_{\rho}$ is the photon
field. In the QCD case, $A_{\rho} = A^{a}_{\rho} T^{a}$,
$A_{\rho}^{a}$ is the non-Abelian gluon field and $T^{a}$ are the
generators of $SU(3)_{c}$ group. 
$"Symm"$ in Eq.(1) means that the limit 
$\epsilon \rightarrow 0$ should be taken symmetrically
\begin{equation}
Symm {\lim_{\epsilon \rightarrow 0 }}\{\frac{\epsilon ^{\mu}}{\epsilon ^2}\}
=0,\ \ \ \ \
Symm {\lim_{\epsilon \rightarrow 0 }}\{\frac{\epsilon ^{\mu}\epsilon ^{\nu}}
{\epsilon ^2}\}=\frac{1}{d}g^{\mu \nu},
\end{equation}
where $d$ denotes the time-space dimension. In this work, we consider four dimensional
gauge field theories, i.e. $d$=4.

Now we compute the curl of the axial-vector current. We have

\begin{eqnarray}
& &\partial_{\lambda}V_5^{\lambda \mu \nu}
=i[\partial^\mu j_5^\nu (x) -\partial^\nu j_5^\mu (x)]
\nonumber \\
&=& Symm {\lim_{\epsilon \rightarrow 0 }}\{\partial_{\lambda}
\bar{\psi}(x+ \epsilon /2)\frac{1}{2} [\gamma ^{\lambda},\sigma^{\mu \nu}]
\gamma_5 U_P(x+ \epsilon /2, x - \epsilon /2) \psi (x - \epsilon /2)\nonumber \\
& &+ \bar{\psi}(x+ \epsilon /2)\frac{1}{2} [\gamma ^{\lambda},\sigma^{\mu \nu}]
\gamma_5 U_P(x+ \epsilon /2, x - \epsilon /2) 
\partial_{\lambda} \psi (x - \epsilon /2)\nonumber \\
& &+ \bar{\psi}(x+ \epsilon /2)\frac{1}{2} [\gamma ^{\lambda},\sigma^{\mu \nu}]
\gamma_5 U_P(x+ \epsilon /2, x - \epsilon /2) 
(-ig \epsilon ^{\rho}\partial_{\lambda}A_{\rho} )
 \psi (x - \epsilon /2)\}.
\end{eqnarray}
To reduce this equation, we use the equations of motion for fermions 
with mass $ m$ : 
\begin{equation}
 (i{\stackrel{\rightarrow}{\makebox[-0.8 mm][l]{/}{D}} }-m)\psi =0,\ \ \ \ \
\bar{\psi}(i\stackrel{\leftarrow}{\makebox[-0.8 mm][l]{/}{D}}+m)=0, 
\end{equation}
where $\vec{D}_\mu =\vec{\partial}_\mu +igA_\mu $ and 
$\stackrel{\leftarrow}{D}_\mu=\stackrel{\leftarrow}{\partial } _\mu -igA_\mu $ 
are coveriant derivatives with 
$A_\mu $ being the gauge fields.
Using Eq.(4), and
keeping terms up to order $\epsilon$, we can reduce Eq.(3) to
\begin{eqnarray}
& &\partial^\mu j_5^\nu (x) -\partial^\nu j_5^\mu (x)
\nonumber \\
&=&{\lim _{x^{\prime }\rightarrow x}}i(\partial _\lambda ^x-\partial _\lambda ^{x^{\prime }})\varepsilon ^{\lambda \mu \nu \rho }\bar{\psi}(x^{\prime })\gamma _\rho U_P (x^{\prime },x)\psi (x) \nonumber \\
& & + Symm {\lim_{\epsilon \rightarrow 0 }}\{ \bar{\psi}(x+ \epsilon /2)
[ - i g (\gamma ^{\nu} \gamma_5 F^{\mu \rho}(x) - \gamma ^{\mu} \gamma_5 F^{\nu \rho}(x) )
\epsilon_{\rho}] \psi(x - \epsilon /2)\} ,
\end{eqnarray}
where $F^{\mu \rho}$ = $\partial^{\mu}A^{\rho} - \partial^{\rho}
A^{\mu}$.

Noticing the fact that the product of the fermion operators is singular,
we must compute the 
singular terms in the operator product of the two fermion fields in the limit 
$\epsilon \rightarrow 0  $. Making the operator product expansion of the two
fermion fields in the presence of a background gauge field, we can perform
the above mentioned computation.
We obtain that the leading term is given by contracting the two operators 
using a free-field propagator, 
which gives zero contribution when trace with  $\gamma^{\mu} \gamma^5$. 
The contribution from second term in the expansion of the product of 
operators leads to[8]
\begin{equation}
\left \langle\bar{\psi}(x+ \epsilon /2)\gamma ^{\mu} \gamma_5  \psi (x - \epsilon /2)\right \rangle
=2g\varepsilon^{\alpha \beta \mu \gamma} F_{\alpha \beta}(x)(\frac{-i}{8\pi^2}\frac{\epsilon_{\gamma}}{\epsilon^2}), 
\end{equation}
where $F_{\alpha \beta}$ = $\partial_{\alpha}A_{\beta} - \partial_{\beta}
A_{\alpha}$. Substituting this expression into Eq.(5) and then taking limit 
in four dimensions, we find 

\begin{eqnarray}
& &\partial ^\mu j_5^\nu (x) -\partial ^\nu j_5^\mu (x)
\nonumber \\
&=&{\lim _{x^{\prime }\rightarrow x}}i(\partial _\lambda ^x-
\partial _\lambda ^{x^{\prime }})\varepsilon ^{\lambda \mu \nu \rho }
\bar{\psi}(x^{\prime })\gamma _\rho U_P (x^{\prime },x)\psi (x) \nonumber \\
& & + \frac{g^2}{16\pi^2} [ \varepsilon ^{\alpha \beta \mu \rho } 
F_{\alpha \beta} (x) F^{\nu}_{\rho}(x) - \varepsilon ^{\alpha \beta \nu \rho }
 F_{\alpha \beta} (x) F^{\mu}_{\rho}(x) ] .
\end{eqnarray}
This is the anomalous axial-vector curl equation, where the last term  
is the anomaly term for the curl of the 
axial-vector current.

The ABJ anomaly, i.e. the axial anomaly, is expressed by the 
anomalous axial-vector divergence equation
\begin{equation}
\partial _\mu j_5^\mu (x)
= - \frac{g^2}{16\pi^2} \varepsilon ^{\alpha \beta \mu \nu } 
F_{\alpha \beta} (x) F_{\mu \nu}(x) 
\end{equation}
for the case of massless fermions. Comparing Eq.(7) and Eq.(8), we find that the anomaly
term for the curl of the axial-vector current is different from the ABJ 
anomaly term, and hence is a type of new one which may be called
the transverse axial anomaly. 

It is well-known that the normal WT relation for the axial-vector vertex is
modified by the axial anomaly.
As a consequence of the transverse axial anomaly, the transverse WT relation
for the axial-vector vertex in
coordinate space (see Eq.(8) of Ref[6]) is also modified as

\begin{eqnarray}
& &\partial _x^\mu \left\langle 0\left| Tj_5^\nu (x)\psi (x_1)\bar{\psi}(x_2)\right| 0\right\rangle -\partial _x^\nu \left\langle 0\left| Tj_5^\mu (x)\psi (x_1)\bar{\psi}(x_2)\right| 0\right\rangle 
\nonumber \\
&=&i\sigma ^{\mu \nu }\gamma _5\left\langle 0\left| T\psi (x_1)\bar{\psi}(x_2)\right| 0\right\rangle \delta ^4(x_1-x)-i\left\langle 0\left| T\psi (x_1)\bar{\psi}(x_2)\right| 0\right\rangle \sigma ^{\mu \nu }\gamma _5\delta ^4(x_2-x) \nonumber \\
& &+{\lim _{x^{\prime }\rightarrow x}}i(\partial _\lambda ^x-\partial _\lambda ^{x^{\prime }})\varepsilon ^{\lambda \mu \nu \rho }\left\langle 0\left| T\bar{\psi}(x^{\prime })\gamma _\rho U_P (x^{\prime },x)\psi (x)\psi (x_1)\bar{\psi}(x_2)
\right| 0\right\rangle \nonumber \\
& & +\frac{g^2}{16\pi^2} \left\langle 0\left| T \psi (x_1)\bar{\psi}(x_2)
[ \varepsilon ^{\alpha \beta \mu \rho } F_{\alpha \beta} (x) F^{\nu}_{\rho}(x) 
- \varepsilon ^{\alpha \beta \nu \rho } F_{\alpha \beta} (x) F^{\mu}_{\rho}(x) ]
\right| 0\right\rangle ,
\end{eqnarray}
where the last term arises from the anomaly term given in Eq.(7).
Accordingly, the transverse WT relation for the axial-vector vertex in  
momentum space (see Eq.(11) of Ref.[6]) is modified as

\begin{eqnarray}
& &iq^\mu \Gamma _A^\nu (p_1,p_2)-iq^\nu \Gamma _A^\mu (p_1,p_2)\nonumber \\
&=&S_F^{-1}(p_1)\sigma ^{\mu \nu }\gamma _5-\sigma ^{\mu \nu }\gamma _5S_F^{-1}(p_2)+(p_{1\lambda }+p_{2\lambda })\varepsilon ^{\lambda \mu \nu \rho }\Gamma _{V\rho }(p_1,p_2) \nonumber \\
& & + \frac {g^2}{ 16 \pi^2} K^{\mu \nu}(p_1,p_2) ,
\end{eqnarray}
where the last term is the anomalous contribution,  and 
$ K^{\mu \nu} $ is defined by
\begin{eqnarray}
& &\int d^{4}x d^{4}x_{1} d^{4}x_{2} e^{i(p_{1}\cdot  x_{1} -
p_{2}\cdot  x_{2} - q\cdot  x)} \langle 0|T \psi(x_{1})
\bar{\psi}(x_{2}) [\varepsilon^{\alpha \beta \mu \rho } F^{\nu}_{\rho }(x) 
- \varepsilon^{\alpha \beta \nu \rho } F^{\mu}_{ \rho }(x) ] F_{\alpha \beta}(x) 
|0\rangle \nonumber \\
&=& (2 \pi)^{4} \delta^{4}(p_{1} - p_{2} - q) iS_{F}(p_{1})
K^{\mu \nu}(p_{1}, p_{2})iS_{F}(p_{2}) .
\end{eqnarray}

Above results are given in the Abelian case. For the axial currents of 
the non-Abelian QCD, the anomaly equation should be the Abelian result, 
supplemented by the appropriate group theory factors[8].
After reading the group theory factors for the anomaly from the fermion loop
diagrams, we find that the curl of flavour 
non-singlet axial currents of
QCD is unaffected by the anomaly of QCD. For the flavour singlet-axial 
current of QCD, the curl of the axial current has the anomaly and 
its form can be obtained from the corresponding result in the case of QED by 
multiplying a factor of $ n_f /2 $, where $ n_f $ is the number of flavors. 
Thus, both longitudinal and transverse WT relations for the flavour 
non-singlet axial-vector vertex and the corresponding vertex function are 
unaffected by the anomaly of QCD. However, both longitudinal and transverse WT 
relations for the flavour-singlet axial-vector vertex and the corresponding 
vertex function have the anomalous contributions. Their forms can be 
written from the corresponding results in QED by multiplying a factor of 
$ n_f /2 $.

By the parallel procedure, we can perform the computation for the curl 
of the vector current operator. We find
\begin{eqnarray}
& &\partial^\mu j^\nu (x) -\partial^\nu j^\mu (x)
\nonumber \\
&=&{\lim _{x^{\prime }\rightarrow x}}i(\partial _\lambda ^x-
\partial _\lambda ^{x^{\prime }})\varepsilon ^{\lambda \mu \nu \rho }
\bar{\psi}(x^{\prime })\gamma _\rho \gamma_5 U_P (x^{\prime },x)\psi (x) \nonumber \\
& & + Symm {\lim_{\epsilon \rightarrow 0 }}\{ \bar{\psi}(x+ \epsilon /2)
[ - i g (\gamma ^{\nu} F^{\mu \rho}(x) - \gamma ^{\mu} F^{\nu \rho}(x) )
\epsilon_{\rho}] \psi(x - \epsilon /2)\} 
\end{eqnarray}
for the case of massless fermions. Applying the method of obtaining 
Eq.(7) to compute the last term of Eq.(12), we find that the contribution of
this term disappears. It shows that there is no anomaly for the vector curl
equation, and hence the transverse WT relation for the vector vertex has
no anomaly.
 
In summary, we have computed the operator equations for the curl of the
axial-vector and the vector currents, respectively, in four dimensional gauge 
theories. In addition to the well-known
anomalous axial-vector divergence equation, we find the anomalous
axial-vector curl equation, which may be called the transverse axial
anomaly. It shows that both normal (longitudinal)
and transverse WT relations for the axial-vector vertex have
anomalies. The computation shows that the transverse WT relation for the 
vector vertex has no
anomaly. As a consequence of anomalies, the full axial-vector
as well as the full vector vertex
functions given by Ref.[6] will be modified in such way that the additional 
terms arising from anomalies
should be added in the vertex funtions[9]. Applying the full vecotr vertex
function including the anomalous contribution to the Dyson-Schwinger
equations for the propagators will imply that the quantum anomaly may play
a significant role in the nonperturbative studies of gauge theories
using Dyson-Schwinger equation formalism, which involves
a deeper aspect of the gauge theories and will be studied further. 

\section*{Acknowledgments}

  This work is supported in part by the National Natural Science Foundation 
of China under Grant No.10075081 and No.19835010.


\begin{thebibliography}{99}
\bibitem{hhx9} Because the vector and the axial-vector vertices are coupled
each other through the transverse WT relations for them (see Ref.[6]),
therefore not only the axial-vector vertex funcion but also the vector vertex
function will be modified by the axial and the transverse axial anomalies.
\end{thebibliography}
\end{document}